\documentclass[11pt,A4paper]{article}
\usepackage{jheppub}
\usepackage{amsmath}
\usepackage{amssymb}
\usepackage{hyperref}
\usepackage{graphicx}

\newcommand{\nn}{\nonumber}

\preprint{OUTP-13-01P}

\title{Soft Supersymmetry Breaking in Anisotropic LARGE Volume Compactifications}

\author{Stephen Angus,}
\author{Joseph P. Conlon}

\affiliation{Rudolph Peierls Centre for Theoretical Physics, University of Oxford, \\
1 Keble Road, Oxford, OX1 3NP, United Kingdom}

\emailAdd{Stephen.Angus@physics.ox.ac.uk}
\emailAdd{Joseph.Conlon@physics.ox.ac.uk}

\abstract{We study soft supersymmetry breaking terms for anisotropic LARGE volume compactifications, where
the bulk volume is set by a fibration with one small four-cycle and one large two-cycle.
We consider scenarios where D7s wrap either a blow-up cycle or the small fibre cycle.
Chiral matter can arise either from modes parallel or perpendicular to the brane. We compute soft terms for this matter and find
that for the case where the D7 brane wraps the fibre cycle the scalar masses
can be parametrically different, allowing a possible splitting of third-generation soft terms.}

\begin{document}
\maketitle
\flushbottom

\section{Introduction}
One of the main goals of string phenomenology is to extract low-energy physics predictions from string theory compactifications.  With the LHC searching for new physics beyond the Standard Model, it is more important than ever to explore the possible TeV-scale particle spectra arising from fundamental theories such as string theory.

Type IIB string theory compactified on Calabi-Yau orientifolds is an attractive framework: the Large Volume Scenario (LVS), in which the extra-dimensional volume is naturally stabilised at values hierarchically larger than the string scale, can give a physical explanation for the large separation that exists between the electroweak scale and the Planck scale \mbox{\cite{0502058, 0505076, 08051029}}.  Meanwhile, moduli-mediated supersymmetry-breaking provides a possible resolution to the hierarchy problem: a light, weakly-coupled Higgs with a mass of around 125GeV is not incompatible with a supersymmetry-breaking scale around the terascale.

The standard realisation of the LARGE Volume Scenario is with an isotropic bulk, where the six extra dimensions are all approximately equally large.
In recent years an interesting variation on the LARGE volume scenario has emerged: anisotropic compactifications, where two of the extra dimensions are stabilised at a scale much larger than the other four \cite{11052107, 11106182, 12036655, 12081160}.
One motivation for this is to attempt to realise the ADD scenario, and work
includes the study of anisotropic large extra dimensions and models of cosmological inflation.

In this paper we take a slightly different tack and study these anisotropic compactifications within the more conventional framework
of supersymmetric solutions to the hierarchy problem.  Our particular aim is to compute soft terms for this scenario:
one motivation for believing this is a worthwhile task is that the anisotropy provides extra geometric structure,
which may in turn generate a non-standard pattern of soft terms. Such a non-standard pattern appears to be necessary
due to the complete absence of any evidence for supersymmetry in current LHC runs.

We consider compactifications whose volume can be expressed in terms of 4-cycles (divisors) as
\begin{equation}
\mathcal{V} = \alpha\Big(\sqrt{\tau_1}\tau_2 - \gamma\tau_3^{3/2}\Big).
\end{equation}
The $\tau_i$ are the real parts of the K\"{a}hler moduli, which determine the sizes of 4-cycles in the extra-dimensional geometry.  The first term corresponds to the bulk volume, while the second term is the
small blow-up cycle that is required in LVS for the bulk volume to be stabilised at large values.  Such structures arise in
K3 fibrations over a $\mathbb{C}P^1$ base; the size of the K3 is given by $\tau_1$.
Examples of explicit geometries of this nature can be found in \cite{11070383, 11103333, 12052485}.
After an appropriate orientifold projection,
the bulk geometric moduli we need to stabilise are the axio-dilaton, the above K\"{a}hler moduli and a number of complex structure moduli (or S-, T- and U-moduli respectively).  The S- and U-moduli are stabilised by background fluxes, while the T-moduli require a combination of $\alpha'$-corrections and non-perturbative effects.

We are interested in anisotropic stabilisation of $\tau_1$ and $\tau_2$, with $\tau_1 \ll \tau_2$. To obtain this we will follow the model
proposed in \cite{11052107}. This model requires poly-instanton contributions \cite{ptsi} to the superpotential from a Euclidean D3 brane wrapping $\tau_1$ and a stack of D7 branes wrapping $\tau_3$.  In the presence of a racetrack superpotential, this setup allows the moduli to be stabilised such that $\tau_1$ and $\tau_3$ are small while $\tau_2$ is exponentially larger \cite{11052107}, generating the required anisotropy.

Poly-instantons are one approach to realising anisotropic geometries --- another approach involving quantum corrections to the K\"ahler potential
is discussed in \cite{11103333}.
There are subtle mathematical questions about the conditions under which such poly-instantons can exist, considered in \cite{12052485}.
We note that the models studied are more string-inspired than string-derived, however
our aim here is not so much to claim any kind of fully honest top-down construction of anisotropic vacua.
To this end we focus on studying the phenomenological consequences,
under the assumption that all the key features of such models can be realised in a consistent manner.

Since the overall volume is exponentially large in the LARGE Volume Scenario, in order to match the Standard Model gauge couplings
it is necessary that the Standard Model arises from branes wrapping a cycle that is either small or collapsed.
We shall consider the cases where the Standard Model gauge groups arise from D7-brane stacks wrapping one of the small cycles in the extra-dimensional geometry.

There are two small 4-cycles that could possibly hold D7s: the small blow-up cycle and the small volume cycle (the K3): we consider each of these cases in turn.  The key determinant of soft terms is the F-terms of the K\"ahler moduli,
\begin{equation}
F^i = \frac{W}{|W|}e^{K/2}K^{i\overline{j}}\big(\overline{W}_{\overline{j}} + \overline{W}K_{\overline{j}}\big),
\end{equation}
as the K\"ahler moduli are dominantly responsible for SUSY breaking.
We also need the matter metric $\tilde{K}_{\overline{\alpha}\beta}$ and its functional dependence on the K\"ahler moduli.
While this cannot be computed directly, we shall argue for its functional form by generalising arguments made in \cite{0609180}.
Matter can arise from modes either interior or transverse to the D7 brane and the form of the K\"ahler metric differs in each case.

The structure of this paper is as follows.
In section 2 we review relevant aspects of the anisotropic Calabi-Yau constructions of \cite{11052107}.  We compute the F-terms for the T-moduli in such constructions in section 3.  Following that, in section 4 we review the generic structure of supersymmetry-breaking soft terms.  Sections 5 and 6 are dedicated to calculating the soft terms for two different scenarios, corresponding to the two possible geometric cycles on which the Standard Model branes could be located.  Finally, we discuss the results and consider the scope for further research.

\section{Anisotropic modulus stabilisation}

In this section we discuss the key features of fibred constructions and how they can lead to anisotropic modulus stabilisation.  We consider one of the simplest scenarios: a K3 fibration over a $\mathbb{C}P^1$ base, with a single del Pezzo divisor localised in the bulk volume.
The fibred structure is essential, as it allows for an anisotropic stabilisation such that the K3 and blow-up mode are small while the $\mathbb{C}P^1$ base is hierarchically larger. While `realistic' compactifications are expected to have hundreds of moduli,
the advantage of the simple model is that it is calculationally tractable.

\subsection{Fibred compactifications}
We start with a Calabi-Yau volume of the form
\begin{equation}
\mathcal{V} = \lambda_1t_1t_2^2 + \lambda_2t_3^3 \; ,
\end{equation}
where the $t_i$ are the volumes of internal 2-cycles in the geometry, while $\lambda_1$ and $\lambda_2$ depend on the particular properties of the Calabi-Yau under consideration. In particular, $t_1$ is the 2-cycle corresponding to the $\mathbb{C}P^1$ base.

The volume can also be expressed in terms of the dual 4-cycles $\tau_i = \partial\mathcal{V}/\partial t_i$,
\begin{equation}
\tau_1 = \lambda_1t_2^2 \; , \;\;\; \tau_2 = 2\lambda_1t_1t_2 \; , \;\;\; \tau_3 = 3\lambda_2\tau_3^2 \; ,
\end{equation}
giving an expression of the form
\begin{equation} \label{eq:vol}
\mathcal{V} = \alpha \Big(\sqrt\tau_1\tau_2 - \gamma\tau_3^{3/2}\Big) = t_1\tau_1 - \alpha\gamma\tau_3^{3/2} \; ,
\end{equation}
where $\alpha = 1/(2\sqrt\lambda_1)$ and $\gamma = \frac{2}{3}\sqrt{\lambda_1/(3\lambda_2)}$.  (In most of what follows we
set $\alpha = \gamma = 1$ for simplicity, unless explicitly stated otherwise.)

\begin{figure}
\includegraphics[width=\textwidth]{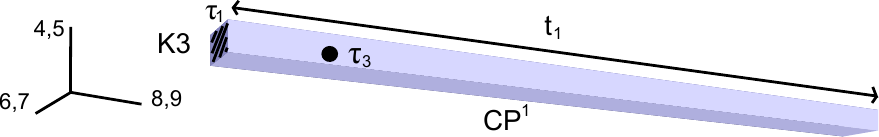}
\caption{Schematic diagram of anisotropic Calabi-Yau structure.} \label{fig:cy}
\end{figure}

In the large-volume regime $t_1\tau_1 \gg \alpha\gamma\tau_3^{3/2}$, so $\mathcal{V} \simeq t_1\tau_1$.  Since $t_1 = (LM_\text{s})^2$ is the size of the $\mathbb{C}P^1$ base (where $M_\text{s}$ is the string scale), $\tau_1 = (lM_\text{s})^4$ is the volume of the K3 fibre.  We are interested in anisotropic compactifications where two of the extra dimensions are hierarchically larger than the other four, which we can achieve if $L\gg l$, or
\begin{equation}
\langle t_1\rangle \gg \sqrt{\langle \tau_1\rangle} \simeq \sqrt{\langle \tau_3 \rangle} \; .
\end{equation}
This corresponds to the $\mathbb{C}P^1$ base being much larger than the K3 fibre, as illustrated in figure~\ref{fig:cy}.

\subsection{Low energy limit}
The stabilisation is described via a four-dimensional supergravity theory.
The K\"ahler moduli present are
\begin{equation}
T_i = \tau_i + i\psi_i \; , \;\; i = 1, \: \dots, h_{1,1} = 3 \; .
\end{equation}
The $\tau_i$ are the volumes of the dual 4-cycles discussed above, while the $\psi_i = \int_{\Sigma_4} C_4$ are their associated axions.
These appear as scalar fields in the effective 4D theory.  The LVS K\"{a}hler potential is
\begin{equation} \label{eq:kah}
K = -2\ln{\bigg( \mathcal{V} + \frac{\xi}{2g_\text{s}^{3/2}}\bigg)} \; ,
\end{equation}
where $g_\text{s}$ is the string coupling and $\xi$ depends on the particular Calabi-Yau in question, but is generally $\sim \mathcal{O}(10^{-2})$.

In order to obtain a compactification that is naturally anisotropic, we follow \cite{11052107} and consider a racetrack superpotential,
\begin{equation}
W = W_0 + Ae^{-a_3T_3} - Be^{-b_3T_3} \; .
\end{equation}
Such a superpotential can arise from gaugino condensation on D7 branes wrapping the blow-up mode $\tau_3$, with the requirement that the gauge group on $\tau_3$ must be chosen to allow condensation into two separate gauge groups.
However, as emphasised earlier, we use this as a phenomenological model and do not claim any kind of top-down derivation.
Poly-instanton corrections may then arise when a Euclidean D3 brane wraps the $\tau_1$ cycle. These give a non-perturbative modification
of the $T_3$ gauge kinetic function,
\begin{equation}
T_3 \rightarrow T_3 + Ce^{-2\pi T_1} \; ,
\end{equation}
which in this scenario ends up being responsible for the anisotropic stabilisation.  The full superpotential then takes the form
\begin{equation} \label{eq:sup}
W = W_0 + Ae^{-a_3(T_3 + C_1e^{-2\pi T_1})} - Be^{-b_3(T_3 + C_2e^{-2\pi T_1})} \; .
\end{equation}
To stay in the semiclassical regime we require that $2\pi\langle\tau_1\rangle \gtrsim 1$, so at leading order
\begin{equation} \label{eq:suplo}
W = W_0 + Ae^{-a_3T_3}\big(1 - a_3C_1e^{-2\pi T_1}\big) - Be^{-b_3T_3}\big(1 - b_3C_2e^{-2\pi T_1}\big) \; .
\end{equation}

The VEVs of the moduli $\tau_i = \text{Re}(T_i)$ and axions $\psi_i = \text{Im}(T_i)$ minimise the supergravity scalar potential
\begin{equation}
V_0 = e^K\Big[K^{i\overline{j}}\big(W_i + WK_i\big)\big(\overline{W}_{\overline{j}} + \overline{W}K_{\overline{j}}\big) -3 |W|^2\Big] \; .
\end{equation}
The minimisation procedure is carried out in two steps.  In the large-volume limit we can expand
\begin{equation}
V_0 = V_{\mathcal{O}(\mathcal{V}^{-3})} + V_{\mathcal{O}(\mathcal{V}^{-3-p})} \; ,
\end{equation}
where $V_{\mathcal{O}(\mathcal{V}^{-3})}$ is the leading-order piece that depends on the overall volume\footnote{In \cite{11052107} it is shown that under reasonable conditions $\psi_3 = 0$ is a minimum of $V_{\mathcal{O}(\mathcal{V}^{-3})}$.  Hence we set $\psi_3 = 0$ in all subsequent calculations.} and $V_{\mathcal{O}(\mathcal{V}^{-3-p})}$ contains the leading $\tau_1$ and $\psi_1$ dependence.  We proceed as follows:
\begin{enumerate}
\item Minimise $V_{\mathcal{O}(\mathcal{V}^{-3})}$ with respect to $\tau_3$ in order to find a useful constraint, which ultimately fixes the VEV of $\tau_3$.
\item Substitute the constraint into $V_{\mathcal{O}(\mathcal{V}^{-3-p})}$ and minimise the result with respect to $\tau_1$ and $\psi_1$.
\end{enumerate}

We focus on the key features of this calculation that are relevant to our purposes.  Writing $a_3 = b_3 + m$, one finds that $V_{\mathcal{O}(\mathcal{V}^{-3})}$ is minimised when
\begin{equation} \label{eq:constr}
e^{-b_3\tau_3} = \frac{3W_0\sqrt{(\tau_3)}}{4Z\mathcal{V}}f_\text{corr} \; ,
\end{equation}
where $Z \equiv Bb_3 - Aa_3e^{-m\tau_3}$ and, to leading order in $\tau_3^{-1}$,
\begin{equation} \label{eq:delta}
f_\text{corr} \,\equiv\, 1 - \frac{3}{4\tau_3}\Bigg[\frac{Z}{Bb_3^2 - Aa_3^2e^{-m\tau_3}}\Bigg] \,\equiv\, 1 - \delta_\text{corr} \; .
\end{equation}
We have assumed that $b_3 \tau_3$ is large enough that $\delta_\text{corr}$ will be sufficiently small for this approximation to be valid.

Substituting (\ref{eq:constr}) into $V_{\mathcal{O}(\mathcal{V}^{-3-p})}$ we eventually find that \cite{11052107}
\begin{equation}
V_{\mathcal{O}(\mathcal{V}^{-3-p})} = \frac{\beta}{\mathcal{V}^3}\big(2\pi\tau_1 - pb_3\tau_3\big)e^{-2\pi\tau_1}\cos{\big(2\pi\psi_1\big)} \; ,
\end{equation}
where
\begin{equation}
\beta \equiv 3W_0^2\sqrt{\tau_3}f_\text{corr}\frac{\big(C_2Bb_3 - C_1Aa_3e^{-m\tau_3}\big)}{Z}
\end{equation}
and
\begin{equation}
p \equiv -\Bigg[\frac{C_2Bb_3^2 - C_1Aa_3^2e^{-m\tau_3}}{C_2Bb_3 - C_1Aa_3e^{-m\tau_3}}\Bigg]\frac{\delta_\text{corr}}{b_3} \; .
\end{equation}
This potential has a global minimum at $\langle\psi_1\rangle = 1/2$ and $2\pi\langle\tau_1\rangle = pb_3\langle\tau_3\rangle + 1$.  (For a more detailed derivation, see \cite{11052107}.)

\subsubsection{A caveat}
From the above calculation it appears that $\langle\tau_1\rangle \propto \langle\tau_3\rangle$, implying that $\tau_1$ is naturally in the correct regime provided $\tau_3$ is small (but sufficiently greater than one that $\delta_\text{corr}$ is small).  Let us now see why this is in fact not the case.  The proportionality cancels because $p \propto \delta_\text{corr} \propto \langle\tau_3\rangle^{-1}$ so
\begin{equation}
pb_3\langle\tau_3\rangle = -\frac{3}{4}\bigg[\frac{C_2Bb_3^2 - C_1Aa_3^2e^{-m\langle\tau_3\rangle}}{C_2Bb_3 - C_1Aa_3e^{-m\langle\tau_3\rangle}}\bigg]\frac{Z}{Bb_3^2 - Aa_3^2e^{-m\langle\tau_3\rangle}} \; .
\end{equation}
This can be rearranged to give
\begin{equation}
pb_3\langle\tau_3\rangle = -\frac{3}{4}\bigg[1 + \frac{(C_2 - C_1)mABa_3b_3e^{-m\langle\tau_3\rangle}}{(C_2Bb_3 - C_1Aa_3e^{-m\langle\tau_3\rangle})(Bb_3^2 - Aa_3^2e^{-m\langle\tau_3\rangle})}\bigg] \; .
\end{equation}
Therefore the VEV of $\tau_1$, corresponding to the volume of the K3 fibration, is given by
\begin{align} \label{eq:tune}
\langle\tau_1\rangle &= \frac{1}{2\pi}\big[pb_3\langle\tau_3\rangle + 1\big] \nn \\ &= \frac{1}{8\pi}\bigg[1 - \frac{3(C_2 - C_1)mABa_3b_3e^{-m\langle\tau_3\rangle}}{(C_2Bb_3 - C_1Aa_3e^{-m\langle\tau_3\rangle})(Bb_3^2 - Aa_3^2e^{-m\langle\tau_3\rangle})}\bigg] \; .
\end{align}
This implies that for natural values $\langle\tau_1\rangle \sim 1/8\pi$, which is beyond the region where the \mbox{$\alpha'$ expansion} can be trusted.

Ideally we would like $\langle\tau_1\rangle$ to be larger; however this requires some fine-tuning.  First of all we require the second term of (\ref{eq:tune}) to be negative, in order to get a positive contribution to $\langle\tau_1\rangle$.  Second, we would like this contribution to be large: the only way to do this is for the denominator to blow up, which is possible if either $C_2Bb_3 \simeq C_1Aa_3e^{-m\langle\tau_3\rangle}$ or alternatively $Bb_3^2 \simeq Aa_3^2e^{-m\langle\tau_3\rangle}$.  Unfortunately the latter would pose problems, since a similar factor appears in the denominator of $\delta_\text{corr}$ --- which we do not want to be too small in case we end up outside the $\delta_\text{corr} \ll 1$ approximation --- so the only option appears to be the former.  One possible solution would be to have $C_2 > C_1$ and $Aa_3e^{-m\langle\tau_3\rangle} > Bb_3$ with $m$ positive.  Then for $1 \gg C_2Bb_3 - C_1Aa_3e^{-m\langle\tau_3\rangle} > 0$ the correction would become large and negative, giving a positive contribution to $\langle\tau_1\rangle$.

We shall proceed on the basis that this issue can be resolved.  One such resolution is that quantum corrections may push $\langle\tau_1\rangle$ into a controlled region.  Alternatively we may hope that, as our interest is in the structural effects of the anisotropy on the soft terms, such structural effects --- in particular the powers of the overall volume that appear --- will be relatively unaffected by a small $\langle\tau_1\rangle$.

\section{F-terms}
We here compute the F-terms,
\begin{equation}
F^i = e^{K/2}K^{i\overline{j}}\big\langle\overline{W}_{\overline{j}} + \overline{W}K_{\overline{j}}\big\rangle \; ,
\end{equation}
where the expectation value is written to emphasise that the VEVs are plugged in after taking derivatives.

Using (\ref{eq:vol}), (\ref{eq:kah}) and (\ref{eq:suplo}) we find that the F-terms are given by
\begin{align}
F^1 &= -\frac{2\langle\tau_1\rangle\tilde{F}}{\mathcal{V} - \frac{\xi}{4g_\text{s}^{3/2}}} \; , \label{eq:f1} \\
F^2 &= -\frac{2\langle\tau_2\rangle\tilde{F}}{\mathcal{V} - \frac{\xi}{4g_\text{s}^{3/2}}} - 4\sqrt{\tau_1}\langle W_1\rangle \; , \label{eq:f2} \\
F^3 &= -\frac{2\langle\tau_3\rangle\tilde{F}}{\mathcal{V} - \frac{\xi}{4g_\text{s}^{3/2}}} + \frac{8\sqrt{\tau_3}}{3}\langle W_3\rangle \; , \label{eq:f3}
\end{align}
where
\begin{align}
\tilde{F} =&\, W_0 + Ae^{-a_3\langle\tau_3\rangle}\Big[\big(1 + 2a_3\langle\tau_3\rangle\big) + a_3C_1e^{-c\langle\tau_1\rangle}\big(1 + 2a_3\langle\tau_3\rangle + 2c\langle\tau_1\rangle\big)\Big] \nonumber \\ &- Be^{-b_3\langle\tau_3\rangle}\Big[\big(1 + 2b_3\langle\tau_3\rangle\big) + b_3C_2e^{-c\langle\tau_1\rangle}\big(1 + 2b_3\langle\tau_3\rangle + 2c\langle\tau_1\rangle\big)\Big] \; .
\end{align}
The leading-order contributions are
\begin{align}
F^1 &= -\frac{2\langle\tau_1\rangle W_0}{\mathcal{V}} \;\; , \\
F^2 &= -\frac{2\langle\tau_2\rangle W_0}{\mathcal{V}} \;\; , \\
F^3 &= -\frac{2\langle\tau_3\rangle W_0\delta_\text{corr}}{\mathcal{V}} \; .
\end{align}

In particular, note that $F^3$ is proportional to $\delta_\text{corr}$ (defined above).  This arises because at leading order $\langle W_3\rangle = (Bb_3e^{-b_3\langle\tau_3\rangle} - Aa_3e^{-a_3\langle\tau_3\rangle}) = e^{-b_3\langle\tau_3\rangle}Z$, so using the condition (\ref{eq:constr}) in (\ref{eq:f3}) we find that
\begin{align}
F^3 &\simeq  -\frac{2\langle\tau_3\rangle W_0}{\mathcal{V}} + \frac{8\sqrt{\tau_3}}{3}\langle W_3\rangle \nonumber \\
&= -\frac{2\langle\tau_3\rangle W_0}{\mathcal{V}}\big(1 - \frac{4e^{-b_3\tau_3}Z\mathcal{V}}{3W_0\sqrt{\tau_3}}\big) \nonumber \\
&= -\frac{2\langle\tau_3\rangle W_0}{\mathcal{V}}\big(1 - f_\text{corr}\big) \; .
\end{align}
Notably, since $\delta_\text{corr} \propto \langle\tau_3\rangle^{-1}$, this means $F^3$ is independent of $\tau_3$ at leading order.  In fact, from the earlier discussion regarding $\langle\tau_1\rangle$, it turns out that $F^1$ and $F^3$ are both simply proportional to $1/\mathcal{V}$, each with a constant of proportionality that depends crucially on the details of the compactification.

At leading order $F^2$ dominates, since we are in the large hierarchy limit where $\langle\tau_2\rangle \gg \langle\tau_1\rangle \sim \langle\tau_3\rangle\delta_\text{corr}$.  Using $\mathcal{V} \simeq \sqrt{\langle\tau_1\rangle}\langle\tau_2\rangle$ we find that
\begin{equation}
F^2 \simeq -\frac{2W_0}{\sqrt{\langle\tau_1\rangle}}
\end{equation}
at leading order.  However, this dominance is not necessarily manifest in the soft terms, to which we now turn.

\section{Soft terms: an overview}
In the following sections we will compute the soft scalar masses, A- and B-terms, as well as the soft gaugino masses, that arise from moduli-mediated supersymmetry-breaking in anisotropic large-volume models.  The moduli responsible for SUSY-breaking are the K\"{a}hler moduli.  We will assume that the Standard Model is present on flux-stabilised D7 branes wrapping one of the small cycles in the extra-dimensional geometry.  Small fluctuations about the vacuum configuration may then give rise to chiral matter.  In this section we review the generic structure of soft terms in the supergravity framework.

The Kahler potential and superpotential can be expanded as a function of observable matter fields $C^\alpha$ to give
\begin{equation} \label{eq:matK}
\mathcal{K} = K(T_i, T_{\overline{i}}^*) + \tilde{K}_{\overline{\alpha}\beta}(T_i, T_{\overline{i}}^*)C^{*\overline{\alpha}}C^\beta + \bigg[\frac{1}{2}Z_{\alpha\beta}(T_i, T_{\overline{i}}^*)C^\alpha C^\beta + \;\text{h.c.}\:\bigg] + \ldots
\end{equation}
and
\begin{equation} \label{eq:matW}
\mathcal{W} = W(T_i) + \frac{1}{2}\mu_{\alpha\beta}C^\alpha C^\beta + \frac{1}{6}Y_{\alpha\beta\gamma}C^\alpha C^\beta C^\gamma + \dots
\end{equation}
respectively, where $K$ and $W$ are the potentials for the moduli only (see (\ref{eq:kah}) and (\ref{eq:sup})).  The convention used is that Greek indices $\alpha$, $\beta$, \ldots run over observable fields while Roman indices $i$, $j$, \ldots correspond to the hidden moduli.  Note that $\mu_{\alpha\beta}$ and $Y_{\alpha\beta\gamma}$ are independent of the $T_i$.  This is because the Peccei--Quinn shift symmetry,
\begin{equation}
\psi_i \rightarrow \psi_i + \epsilon_i \; ,
\end{equation}
prevents the K\"{a}hler moduli from appearing in the holomorphic function $\mathcal{W}$ (except as non-perturbative terms, e.g. (\ref{eq:sup})).  Since the superpotential is not renormalised at any order in perturbation theory, the symmetry is unbroken perturbatively.  Hence the holomorphic terms are functions of the axio-dilaton and complex structure moduli only, however in the present scenario we have integrated those out.  We conclude that the only remaining places where K\"{a}hler moduli can appear are in the K\"{a}hler matter metric $\tilde{K}_{\overline{\alpha}\beta}$ and the function $Z_{\alpha\beta}$.  In general these are highly non-trivial to compute.  However it turns out that, in the cases we want to consider, the dependence of $\tilde{K}_{\overline{\alpha}\beta}$ on the T-moduli can be deduced through scaling arguments.  Finally, in order to fix $Z_{\alpha\beta}$ we assume that its dependence on the K\"ahler moduli is related to that of $\tilde{K}_{\overline{\alpha}\beta}$, as would be the case if these terms were to share a common origin in the fundamental theory.

For now, let us compute the general structure of soft terms.  By plugging $\mathcal{K}$ and $\mathcal{W}$ into the Supergravity scalar potential,\footnote{The convention here is that indices $I$, $J$, \ldots run over both hidden and observable fields.}
\begin{equation}
V = e^\mathcal{K}\bigg[\mathcal{K}^{I\overline{J}}\big(\mathcal{W}_I + \mathcal{W}\mathcal{K}_I\big)\big(\overline{\mathcal{W}}_{\overline{J}} + \overline{\mathcal{W}}\mathcal{K}_{\overline{J}}\big) -3 |\mathcal{W}|^2\bigg] \; ,
\end{equation}
and taking the limit $M_P \rightarrow \infty$ to neglect non-renormalisable (hard) terms, we find that the scalar potential becomes
\begin{equation}
V = V_0 + V_\text{soft} \; .
\end{equation}
The soft scalar potential $V_\text{soft}$ can be written in the form
\begin{equation}
V_\text{soft} = \big(m_0^2 + m'^2\big)_{\overline{\alpha}\beta}C^{*\overline{\alpha}}C^\beta + \bigg(\frac{1}{2}B'_{\alpha\beta}C^\alpha C^\beta + \frac{1}{6}A'_{\alpha\beta\gamma}C^\alpha C^\beta C^\gamma + \;\text{h.c.}\:\bigg) \; ,
\end{equation}
where $(m_0^2)_{\overline{\alpha}\beta}$ is a supersymmetric mass term,\footnote{To be precise, $(m_0^2)_{\overline{\alpha}\beta} = \mu'_{\alpha\beta}\tilde{K}^{\alpha\overline{\beta}}\overline{\mu'}_{\overline{\beta}\overline{\alpha}}$, where $\mu'_{\alpha\beta}$ is the effective $\mu$ parameter of (\ref{eq:hatmu}).} while $m'^2_{\overline{\alpha}\beta}$, $A'_{\alpha\beta\gamma}$ and $B'_{\alpha\beta}$ are the un-normalised scalar masses, trilinear and bilinear terms respectively:
\begin{align}
m'^2_{\overline{\alpha}\beta} \; = \; &\Big(m_{3/2}^2 + V_0\Big)\tilde{K}_{\overline{\alpha}\beta} \nn \\
&- \overline{F}^{\overline{i}} \Big(\partial_{\overline{i}}\partial_j\tilde{K}_{\overline{\alpha}\beta} - \partial_{\overline{i}}\tilde{K}_{\overline{\alpha}\gamma}\tilde{K}^{\gamma\overline{\delta}}\partial_j\tilde{K}_{\overline{\delta}\beta}\Big)F^j \; ; \\
A'_{\alpha\beta\gamma} \; = \; &\frac{\overline{W}}{|W|}e^{K/2}F^i \Big[K_i Y_{\alpha\beta\gamma} + \partial_i Y_{\alpha\beta\gamma} \nn \\
&- \Big(\tilde{K}^{\delta\overline{\rho}}\partial_i\tilde{K}_{\overline{\rho}\alpha}Y_{\delta\beta\gamma} + (\alpha \leftrightarrow \beta) + (\alpha \leftrightarrow \gamma)\Big)\Big] \; ;
\end{align}
\begin{align}
B'_{\alpha\beta} \; = \; &\frac{\overline{W}}{|W|}e^{K/2} \Big\{F^i \Big[K_i \mu_{\alpha\beta} + \partial_i \mu_{\alpha\beta} \nn \\
&- \Big(\tilde{K}^{\delta\overline{\rho}}\partial_i \tilde{K}_{\overline{\rho}\alpha}\mu_{\delta\beta} + (\alpha \leftrightarrow \beta)\Big)\Big] - m_{3/2}\mu_{\alpha\beta}\Big\} \nn \\
&+ \Big(2m_{3/2}^2 + V_0\Big) Z_{\alpha\beta} - m_{3/2}\overline{F}^{\overline{i}}\partial_{\overline{i}}Z_{\alpha\beta} \nn \\
&+ m_{3/2}F^i \Big[\partial_i Z_{\alpha\beta} - \Big(\tilde{K}^{\delta\overline{\rho}}\partial_i \tilde{K}_{\overline{\rho}\alpha}Z_{\delta\beta} + (\alpha \leftrightarrow \beta)\Big)\Big] \nn \\
&- \overline{F}^{\overline{i}}F^j \Big[\partial_{\overline{i}}\partial_j Z_{\alpha\beta} - \Big(\tilde{K}^{\delta\overline{\rho}}\partial_j \tilde{K}_{\overline{\rho}\alpha}\partial_{\overline{i}}Z_{\delta\beta} + (\alpha \leftrightarrow \beta)\Big)\Big] \; .
\end{align}

For a diagonal K\"{a}hler matter metric,
\begin{equation}
\tilde{K}_{\overline{\alpha}\beta} = \tilde{K}_\alpha\delta_{\overline{\alpha}\beta} \; ,
\end{equation}
we can simplify many of the soft term expressions.  The B-term is only relevant for the Higgs fields, for which we require that only $Z_{H_1H_2} = Z_{H_2H_1} \equiv Z$ is non-zero.  We expect the superpotential $\mu$ term to vanish for all soft matter, since its magnitude is set by the Planck scale and a non-zero value would lift the masses up to that scale.  Nevertheless, we define $\mu_{H_1H_2} = \mu_{H_2H_1} \equiv \mu$ in order to better understand the structure of the B-term calculation, before finally taking the limit $\mu \rightarrow 0$.  Hence the complete soft term Lagrangian, including gaugino mass terms, becomes \cite{ssb}
\begin{align}
\mathcal{L}_\text{soft} = \; &\frac{1}{2}(M_a\widehat{\lambda}^a\widehat{\lambda}^a + \text{h.c.}) -
m_\alpha^2\widehat{C}^{*\overline{\alpha}}\widehat{C}^\alpha \nn \\ &- \bigg(\frac{1}{6}A_{\alpha\beta\gamma}\widehat{Y}_{\alpha\beta\gamma}\widehat{C}^\alpha\widehat{C}^\beta\widehat{C}^\gamma + B\widehat{\mu}\widehat{H}_1\widehat{H}_2 + \text{h.c.}\bigg) \; ,
\end{align}
where it is now convenient to use canonically normalised soft matter fields, e.g. scalar fields,
\begin{align}
\widehat{C}^\alpha &= \tilde{K}_\alpha^{1/2}C^\alpha \; , \; \\
\intertext{and gauginos,}
\widehat{\lambda}^a &= (\text{Re}(f_a))^{1/2}\lambda^a \; .
\end{align}
Here $f_a$ is the gauge kinetic function (the index $a$ runs over gauge groups).  We have also introduced the physical Yukawa couplings,
\begin{align}
\widehat{Y}_{\alpha\beta\gamma} \; = \; &Y_{\alpha\beta\gamma}\frac{\overline{W}}{|W|}e^{K/2}(\tilde{K}_\alpha\tilde{K}_\beta\tilde{K}_\gamma)^{-1/2} \; , \\
\intertext{and the rescaled $\mu$ parameter,}
\widehat{\mu} \; \equiv \; &(\tilde{K}_{H_1}\tilde{K}_{H_2})^{-1/2}\mu' \; , \\
\intertext{where}
\mu' \; \equiv \; &\frac{\overline{W}}{|W|}e^{K/2}\mu + m_{3/2}Z - \overline{F}^{\overline{i}}\partial_{\overline{i}} Z \; . \label{eq:hatmu}
\end{align}

With these simplifying assumptions, the soft terms are given by the expressions
\begin{align} \label{eq:Mg}
M_a &= \frac{1}{2\text{Re}(f_a)}F^m\partial_m f_a \; , \\ \label{eq:m2}
m_\alpha^2 \; &= \; \Big(m_{3/2}^2 + V_0\Big) - \overline{F}^{\overline{i}}F^j \partial_{\overline{i}}\partial_j\ln\tilde{K}_\alpha \; , \\ \label{eq:A}
A_{\alpha\beta\gamma} \; &= \; F^i \Big[K_i + \partial_i \ln Y_{\alpha\beta\gamma} - \partial_i \ln(\tilde{K}_\alpha\tilde{K}_\beta\tilde{K}_\gamma)\Big] \; , \\
B \; &= \; \widehat{\mu}^{-1}(\tilde{K}_{H_1}\tilde{K}_{H_2})^{-1/2}\bigg\{\frac{\overline{W}}{|W|}e^{K/2}\mu\Big(F^i \Big[K_i + \partial_i\ln\mu \nn \\
&\;\;\;\;\, - \partial_i \ln(\tilde{K}_{H_1}\tilde{K}_{H_2})\Big] - m_{3/2}\Big) \nn \\
&\;\;\;\;\, + \Big(2m_{3/2}^2 + V_0\Big)Z - m_{3/2}\overline{F}^{\overline{i}}\partial_{\overline{i}} Z \nn \\
&\;\;\;\;\, + m_{3/2}F^i \Big[\partial_i Z - Z\partial_i \ln(\tilde{K}_{H_1}\tilde{K}_{H_2})\Big] \nn \\ \label{eq:B}
&\;\;\;\;\, - \overline{F}^{\overline{i}}F^j \Big[\partial_{\overline{i}}\partial_j Z - \partial_{\overline{i}}Z\partial_j\ln(\tilde{K}_{H_1}\tilde{K}_{H_2})\Big]\bigg\} \; .
\end{align}

In the large hierarchy model there are two possible cycles on which chiral matter could be located: the blow-up mode $\tau_3$ and the small cycle $\tau_1$ corresponding to the K3 fibration. The large cycle $\tau_2$ is ruled out based on the observed values of the Standard Model
gauge couplings.
For each scenario we will first need to deduce the K\"{a}hler matter metric $\tilde{K}_{\overline{\alpha}\beta}$ in order to be able to compute soft terms.  We consider each case in turn.

\section{D7s wrapping the blow-up mode \boldmath $\tau_3$} \label{sec:t3}
The first possibility is that the Standard Model arises from magnetised D7 branes localised on the blow-up mode of size $\tau_3$.
As non-perturbative effects are located on this cycle, there may be a
tension between the chiral nature of the Standard Model and the non-perturbative effects \cite{07113389}, although see \cite{11053193} for arguments that this can be evaded.

To determine the K\"{a}hler matter metric we use an argument articulated in \cite{0609180}. We first deduce how $\tilde{K}_{\overline{\alpha}\beta}$ depends on the volume cycles $\tau_1$ and $\tau_2$ by assuming that the physical Yukawa couplings $\hat{Y}_{\alpha\beta\gamma}$ are independent of the bulk volume $\mathcal{V}$ and subsequently use scaling arguments to extract the leading-order $\tau_3$ dependence. We then compute the soft terms and find that they are all of order $m_{3/2}$ multiplied by a universal factor $\delta_\text{corr}$, which depends on the details of the compactification.

\subsection{Computing the K\"{a}hler matter metric}
We can deduce the K\"{a}hler matter metric $\tilde{K}_{\overline{\alpha}\beta}$ by considering the canonical normalisation of the Yukawa couplings,\footnote{We have assumed a diagonal K\"{a}hler matter metric for simplicity, $\tilde{K}_{\overline{\alpha}\beta} = \tilde{K}_\alpha\delta_{\overline{\alpha}\beta}$, since the results here do not depend on the structure of $\tilde{K}_{\overline{\alpha}\beta}$.}
\begin{equation} \label{eq:yuk}
\widehat{Y}_{\alpha\beta\gamma} = \frac{e^{K/2}Y_{\alpha\beta\gamma}}{\big(\tilde{K}_\alpha \tilde{K}_\beta \tilde{K}_\gamma\big)^{1/2}} \; .
\end{equation}
Here the $\widehat{Y}_{\alpha\beta\gamma}$ are physical Yukawa couplings while the $Y_{\alpha\beta\gamma}$ are the corresponding superpotential Yukawa couplings (\ref{eq:matW}). The $Y_{\alpha\beta\gamma}$ are independent of the K\"{a}hler moduli, while the $\tau_i$-dependence of $K$ is known.

Let us now consider the $T_i$-dependence of the physical Yukawa couplings $\widehat{Y}_{\alpha\beta\gamma}$.  These are marginal couplings
generated by local interactions within the bulk, so we want them to remain finite when we decouple gravity, i.e. when we take the limit $M_\text{P} \rightarrow \infty$.  In addition, as matter fields are localised on the blow-up cycle they should be unaffected by a rescaling of the bulk volume $\mathcal{V}$.  Therefore we deduce that the physical Yukawas must be independent of $\mathcal{V} \simeq \sqrt{\tau_1}\tau_2$ at leading order in the large-volume expansion.

Furthermore, even though the bulk is anisotropic, by assumption all cycles are large enough to be outside the quantum regime.  Since in this case the Yukawa couplings are localised at a singularity, this ensures that they should also be independent of anisotropic rescalings of the bulk cycles.  This then implies that the physical Yukawas are independent of \mbox{both $\tau_1$ and $\tau_2$}.

Therefore in the present scenario we expect the leading contribution to the matter metric to be
\begin{equation} \label{eq:Kmm1}
\tilde{K}_\alpha = \frac{k_\alpha'}{\mathcal{V}^{2/3}} = \frac{k_\alpha'}{(\tau_1\tau_2^2)^{1/3}} \; .
\end{equation}
Note that this result contains the same $\mathcal{V}$-dependence as the simpler case where a single large cycle controls the bulk volume \cite{0609180}.

Now we turn to the dependence of the matter metric on the small blow-up cycle $\tau_3$.  Assuming we are in the geometric regime,
the leading-order T-moduli dependence of the K\"{a}hler matter metric is given by
\begin{equation} \label{eq:kt3}
\tilde{K}_{\overline{\alpha}\beta} = \frac{\tau_3^{\lambda_\alpha}}{\left( \tau_1\tau_2^2 \right)^\frac{1}{3}}k_{\overline{\alpha}\beta} \; .
\end{equation}
The function $k_{\overline{\alpha}\beta}$ depends only on S- and U-moduli, so we treat it as a constant.
The value of $\lambda$ depends on whether matter originates as bulk matter, with support across the whole 4-cycle, or as a `matter curve', with
support only on a 2-dimensional subspace of the local 4-cycle.

\subsection{Soft terms}

The soft terms in this case are very similar to an analogous calculation in \cite{0610129}.
The gauge kinetic function $f_a$ is simply
\begin{equation}
f_a = k_a T_3 \; ,
\end{equation}
for $k_a$ an appropriate constant. The gaugino mass (\ref{eq:Mg}) is then
\begin{equation} \label{eq:ggino}
M_a = -\frac{\delta_\text{corr} W_0}{\mathcal{V}} \; , \;\;\;\; \forall \; a \; ,
\end{equation}
where $\delta_\text{corr}$ is defined in (\ref{eq:delta}).

The soft scalar masses and trilinear A-terms are given by (\ref{eq:m2}) and (\ref{eq:A}) respectively.  These are:
\begin{equation}
m_\alpha^2 = \frac{\lambda_\alpha(\delta_\text{corr})^2 W_0^2}{\mathcal{V}^2} \; ; \;\;\;\; A = \frac{\delta_\text{corr} W_0}{\mathcal{V}} \; .
\vspace{0.25cm}
\end{equation}
The A-terms are universal due to the constraint $\lambda_\alpha + \lambda_\beta + \lambda_\gamma = 1$, which is required in order to get the correct scaling for the Yukawa couplings.

Finally, we turn to the B-term.  We do not specify the geometric origin of the Higgs doublets, so for generality we express the scaling of their K\"{a}hler matter metric components as
\begin{equation}
\tilde{K}_{H_i} = \frac{\tau_3^{\lambda_{H_i}}}{\left( \tau_1\tau_2^2 \right)^\frac{1}{3}}k_{H_i} \; , \;\; i = 1,2 \; .
\end{equation}
The function $Z(T_i, T_{\overline{i}}^*)$ is in general unknown and hard to compute, since it is not protected by holomorphy.  However, we can proceed by making the assumption that the scaling of $Z$ with $\tau_3$ is related to the scaling of $\tilde{K}_{H_1}$ and $\tilde{K}_{H_2}$ with $\tau_3$, which would be the case if these terms all had the same origin in the fundamental theory.  Using the fact that $Z \equiv Z_{H_1 H_2}$ and interpreting $Z$, $\tilde{K}_{H_1}$ and $\tilde{K}_{H_2}$ as products and squares of vielbeins, one can see that $Z$ should scale as $\sqrt{\tilde{K}_{H_1}\tilde{K}_{H_2}}$.  Therefore
\begin{equation}
Z = \frac{\tau_3^{\overline{\lambda}}}{\left( \tau_1\tau_2^2 \right)^\frac{1}{3}}z \; ,
\end{equation}
where $z$ is independent of the K\"{a}hler moduli and $\overline{\lambda} \equiv (\lambda_{H_1} + \lambda_{H_2})/2$.

Using this information, and setting the superpotential $\mu = 0$, one can use (\ref{eq:B}) to calculate the B-term.  The result is
\begin{equation}
B = \frac{(\overline{\lambda} + 1)\delta_\text{corr}W_0}{\mathcal{V}} \; .
\end{equation}

The key feature here is that the soft terms are all of the same order and all comparable to the gravitino mass.  Note that they are all multiplied by a factor $\delta_\text{corr}$, which is inversely proportional to $\langle\tau_3\rangle$.  The no-scale structure is broken by $F^3$, the F-term corresponding to the blow-up mode, as one would expect based on simpler large-volume models.

\section{D7s wrapping the small volume cycle \boldmath $\tau_1$}

We now consider the realisation of the Standard Model by wrapping D7 branes on the small volume cycle of size $\tau_1$ (corresponding to the K3 fibre).  We assume that this can be done consistently with the generation of appropriate structures to realise the anisotropic stabilisation, in concordance with our aim of exploring the possible soft-term structures that can arise from anisotropic constructions.

Under this assumption, we compute the soft terms for chiral matter on the D7s wrapping $\tau_1$.  Two types of matter are possible: modes $\phi$ corresponding to the position of the D7 stack in transverse space; and `longitudinal' modes, coming from the massless modes of 8-dimensional gauge multiplet fields $A$ inside the D7 worldvolume \cite{0805.2943}.  It turns out that the anisotropy naturally generates a large hierarchy between generations of soft terms.  Again we begin by computing the T-moduli dependence of the K\"{a}hler matter metric for the two types of matter.  To this end, we consider first the result for a 6-torus (projected as $T^2 \times T^2 \times T^2$).  We then deduce from the expression for the volume $\mathcal{V}$ how the result is modified in the present scenario.  Finally we compute soft terms.

\subsection{K\"{a}hler matter metric with two components}
\begin{figure}
\includegraphics[width=\textwidth]{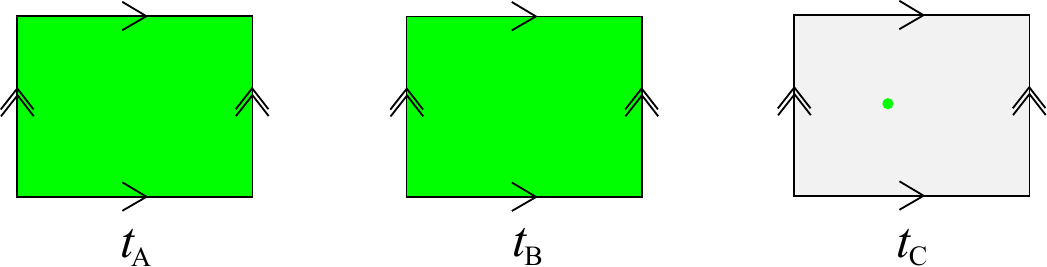}
\caption{Compactification on $T^2 \times T^2 \times T^2$ with D7 branes wrapping tori A and B.  The D7s are pointlike in torus C and free to move about.} \label{fig:T2}
\end{figure}

First, a toroidal example.  The bulk volume is simply given by a product of 2-cycles,
\begin{equation}
\mathcal{V} = t_\text{A} t_\text{B} t_\text{C} \; ,
\end{equation}
where the 2-cycles correspond to areas of tori, labelled A, B and C.  It turns out that, for the case of D7 branes transverse to the $t_\text{C}$ direction (see figure~\ref{fig:T2}), the transverse and internal components of the K\"{a}hler matter metric are (e.g. see \cite{0805.2943})
\begin{equation}
\tilde{K}_{\parallel} = \tilde{K}_{(7^\text{C}7^\text{C})_\text{B}} = \frac{1}{(2\tau_\text{A})} \; ; \;\;\; \tilde{K}_{\perp} = \tilde{K}_{(7^\text{C}7^\text{C})_\text{C}} = \frac{g_\text{s}}{2} \; ,
\end{equation}
where $\tau_\text{A}$ is the dual 4-cycle to $t_\text{A}$.  The $7^\text{C}$ refers to a D7 brane transverse to the complex plane of torus C, while the outer subscript indicates in which plane the string modes exist.  For example, the first term refers to string modes inside the D7 worldvolume, which at low energies correspond to components $A$ of brane-worldvolume fluxes.  The second term corresponds to tranverse string modes, which are related to the position of the stack of D7s on torus C.  At low energies they are realised as scalars $\phi$ in the effective 8D theory on the brane.  These components of the matter metric turn out to be independent of the T-moduli.

\begin{figure}
\includegraphics[width=\textwidth]{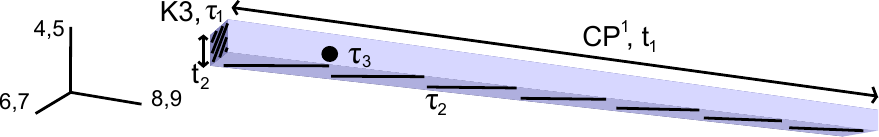}
\caption{Geometrical meaning of $t_1$, $t_2$, $\tau_1$ and $\tau_2$.} \label{fig:cy2}
\end{figure}

Now let us compare the above simple scenario to our model.  Neglecting the blow-up mode, the volume is given by
\begin{equation}
\mathcal{V} \simeq t_1\tau_1 \simeq t_1t_2^2
\end{equation}
so the obvious schematic relations between the 2-cycles and their dual 4-cycles are $\tau_1 \sim t_2^2$ and $\tau_2 \sim t_1t_2$ (see figure~\ref{fig:cy2}).  Since the D7 branes wrap $\tau_1$ they are transverse to $t_1$, so this 2-cycle plays the same role as $t_\text{C}$ above.  The role of $t_2$ is slightly more subtle, but it essentially plays the role of $t_\text{A} = t_\text{B}$.  Since $\tau_1$ corresponds to the K3, $t_2$ is effectively the `square root' of the K3 volume.

From this discussion, we conclude that the components of the K\"{a}hler matter metric are given by
\begin{equation}
\label{eqeq}
\tilde{K}_{\parallel} = \frac{1}{(2\tau_2)} \; ; \;\;\; \tilde{K}_{\perp} = \frac{g_\text{s}}{2} \; .
\end{equation}
Matter in a single generation --- those fields with identical gauge charges --- can have distinct geometric origins, and thereby distinct
K\"ahler metrics. The different volume scalings of these K\"ahler metrics can lead to different soft terms, as we shall now discover.

\subsection{Soft terms revisited}
We can now compute soft terms for the scenario in which Standard Model D7s wrap the small volume cycle $\tau_1$.
The gaugino masses are once again given by (\ref{eq:Mg}),
\begin{equation}
M_a = \frac{1}{2\text{Re}(f_a)}F^m\partial_mf_a \; ,
\end{equation}
but this time the gauge kinetic function, $f_a = k_a T_1$.  Therefore,
\begin{equation}
M_a = -\frac{W_0}{\mathcal{V}} \; , \;\;\;\; \forall \; a \; ,
\end{equation}
i.e. the gaugino masses have the same magnitude as the gravitino mass.

There are now two possible soft scalar masses, corresponding to transverse ($\phi$) and internal D7 worldvolume ($A$) modes.
The different matter metrics in eq. (\ref{eqeq}) give different soft terms,
\begin{equation}
m_\perp^2 = \frac{W_0^2}{\mathcal{V}^2} + \mathcal{O}(\mathcal{V}^{-3}) \; , \;\;\; m_\parallel^2 = \mathcal{O}(\mathcal{V}^{-3}) \; .
\end{equation}
We find that the transverse scalars have masses of order the gravitino mass, \mbox{$m_{3/2} = W_0/\mathcal{V}$}, while the internal scalars are suppressed by a factor of $1/\mathcal{V}$.

Next we turn to the A-terms.  There are now four possibilities, depending on which of the three interacting scalars are transverse or internal.  These turn out to be
\begin{align} \nn
A_{\perp\perp\perp} &= \frac{3W_0}{\mathcal{V}} \; , &A_{\perp\perp\parallel} &= \frac{2W_0}{\mathcal{V}} \; , \\ A_{\perp\parallel\parallel} &= \frac{W_0}{\mathcal{V}} \; , \; &A_{\parallel\parallel\parallel} &= \mathcal{O}(\mathcal{V}^{-2}) \; .
\end{align}
Note that the A-terms are proportional (at tree-level) to the integer number of transverse $\phi$ matter modes involved in the interaction, so the trilinear term corresponding to the A-A-A interaction is strongly suppressed.

Finally, we compute the B-term.  To this end, it is worth recalling how the B-term appears in the MSSM Lagrangian:
\begin{equation}
\mathcal{L}_B = -\Big(B\widehat{\mu}\widehat{H_1}\widehat{H_2} \; + \; \text{h.c.}\Big) \; ,
\end{equation}
where $\widehat{H_1}$ and $\widehat{H_2}$ are the Higgs doublets.  Note in particular that the bilinear term scales as the combination $B\widehat{\mu}$, so it is possible for $B$ to diverge at leading order while the overall bilinear term remains finite.  To acknowledge this point, we carry out the calculation for arbitrary $\mu$ and take the limit $\mu \rightarrow 0$.

There are three different possible values for the B-term, depending on the geometric origin of each Higgs doublet.  These correspond respectively to scenarios where: both doublets arise from transverse $\phi$ scalars; one doublet is a transverse $\phi$ mode and the other is an internal $A$ mode; or both Higgs doublets are internal $A$ modes.  Since the K\"{a}hler matter metric only depends on $\tau_2$ we restrict our focus to this modulus.  As in section \ref{sec:t3}, we assume that the power dependence of $Z$ on the relevant modulus is the mean of the dependences of $\tilde{K}_{H_1}$ and $\tilde{K}_{H_2}$.  For $\tilde{K}_{H_1} \sim \tau_2^{-\lambda_{H_1}}$, $\tilde{K}_{H_2} \sim \tau_2^{-\lambda_{H_2}}$ and $Z \sim \tau_2^{-\overline{\lambda}}$, where $\overline{\lambda} = (\lambda_{H_1} + \lambda_{H_2})/2$, we find that
\begin{equation}
B = \frac{2m_{3/2}\big(1 - \overline{\lambda}\big)\big\{e^{K/2}\mu + \big(1 - \frac{1}{2}\overline{\lambda}\big)m_{3/2}Z\big\}}{\mu'} \; ,
\end{equation}
where
\begin{equation} \label{eq:mup}
\mu' \equiv \big(\tilde{K}_{H_1}\tilde{K}_{H_2}\big)^\frac{1}{2}\widehat{\mu} = e^{K/2}\mu + \big(1 - \overline{\lambda}\big)m_{3/2}Z \; .
\end{equation}
We consider each possible value of $\overline{\lambda}$ and evaluate the respective B-term.

\paragraph{1. Both doublets arise from transverse $\phi$ scalars.} $\;$ \\
Here $\tilde{K}_{H_1}$, $\tilde{K}_{H_2}$ and $Z$ are all independent of $\tau_2$, so $\overline{\lambda} = 0$. We find that
\begin{equation}
B = 2m_{3/2} = \frac{2W_0}{\mathcal{V}} \; .
\end{equation}
This result holds regardless of the value of the superpotential $\mu$ term.

\paragraph{2. One doublet is a transverse $\phi$-mode and the other is an internal $A$-mode.} $\;$ \\
We now have $\tilde{K}_{H_1} \sim \tau_2^{-1}$ and $\tilde{K}_{H_2} \sim \tau_2^{0}$ (or vice-versa), so $Z \sim \tau_2^{-1/2}$.  Therefore,
\begin{equation}
B = \frac{3m_{3/2}}{2} = \frac{3W_0}{2\mathcal{V}} \; .
\end{equation}

\paragraph{3. Both Higgs doublets are internal $A$-modes.} $\;$ \\
Finally, we consider the scenario where $\tilde{K}_{H_1} \sim \tilde{K}_{H_2} \sim Z \sim \tau_2^{-1}$.  In this case $\overline{\lambda} = 1$, which implies that $B$ is undefined: after taking the limit $\mu \rightarrow 0$ we find that $\widehat{\mu} = 0$ at leading order.  Hence the denominator of $B$, given by (\ref{eq:mup}), vanishes; however the physical B-term depends on the combination $B\widehat{\mu}$.  We find that for $\overline{\lambda} = 1$ the numerator of $B$ also vanishes, so
\begin{equation}
B\widehat{\mu} = 0
\end{equation}
at leading order in $1/\mathcal{V}$.

For our purposes the latter two possibilities are more interesting, since they involve Higgs modes that are naturally suppressed with respect to the gravitino mass.  In the final scenario the B-term itself is suppressed.

\subsection{Low-energy consequences}

At first glance, one would be tempted to conclude that there can be an inter-generational soft term splitting of order $1/\mathcal{V}$.
This is interesting because various models of so-called natural supersymmetry rely on light third-generation soft terms with heavier scalar masses
for the first two generations (e.g. see \cite{0602096}, and for a stringy model \cite{12014857}).

 However, the soft terms have been evaluated at tree-level and at the compactification scale.  To obtain the soft terms observed at TeV-scale, one must integrate out the higher-energy modes via the renormalisation group flow, and in doing so include loop corrections.
Such radiative corrections will tend to reduce the inter-generational splitting, and as all soft terms feed into one another we expect the low energy
splitting to be no larger than a loop factor. This scenario requires $\mathcal{V} \sim \mathcal{O}(10^{14})$, so we would have a UV string scale of order $m_\text{s} \sim 10^{11}\,\text{GeV}$. As this is expected to be the UV scale for soft term and gauge coupling running, such a scenario would not be compatible with any kind of conventional grand unification.

\section{Conclusions}
In this paper we have computed soft terms for anisotropic large-volume Calabi-Yau compactifications, assuming that the chiral matter of the Standard Model is located on flux-stabilised D7 branes wrapping one of the small cycles.  The anisotropic models we have considered have a volume of the form
\begin{equation}
\mathcal{V} = \alpha \Big(\sqrt\tau_1\tau_2 - \gamma\tau_3^{3/2}\Big)
\end{equation}
when expressed in terms of the real parts of K\"{a}hler moduli, which correspond to the sizes of 4-cycles in the geometry.  Two of these moduli correspond to small cycles: the blow-up cycle $\tau_3$ which is localised in the bulk, and the small volume cycle $\tau_1$.  We have considered what happens when a stack of D7s wraps each of these small cycles and computed the associated soft terms.

When the chiral matter of the Standard Model is produced by magnetised D7 branes wrapping the blow-up mode, we find soft terms that are of order $m_{3/2}$ and multiplied by a universal factor that depends upon the details of the compactification.  This is a typical structure of the kind one would expect based on similar, simpler large-volume models.  On the other hand, when the Standard Model comes from additional D7s wrapping the small volume cycle there is a splitting between generations of soft terms, which is a new feature.  This can be understood heuristically as coming directly from the anisotropy, since some modes are aligned along the large directions transverse to the D7 worldvolume, while other internal D7 modes oscillate along the small cycle directions.  Some of the soft terms are of order $m_{3/2}$, while others (those corresponding to modes in the D7 worldvolume) are suppressed at tree-level by a factor of $1/\mathcal{V}$.

For the case of D7s wrapping $\tau_1$, we compared the Calabi-Yau structure with the toroidal case and constructed the matter metric by analogy.  We found two different terms, depending on the higher-dimensional origin of matter:
\begin{equation}
\tilde{K}_{\parallel} = \frac{1}{(2\tau_2)} \; ; \;\;\; \tilde{K}_{\perp} = \frac{g_\text{s}}{2} \; .
\end{equation}
Note that one of these is independent of the K\"{a}hler moduli, while the other is not.  The suppression of the soft terms corresponding to D7 worldvolume oscillations is a direct consequence of this fact.

Let us finally mention some limitations of our results. Our interest has been in the phenomenology of anisotropy and we have
simply assumed the validity of the poly-instanton approach to constructing an anisotropic
compactification. It is fair to say that such approaches are at best string-inspired rather than string-derived.
Furthermore, to generate the splitting in soft terms one needs to wrap an extra stack of D7 branes around the K3.  It is conceivable that instanton corrections generated by these D7s could dominate over the poly-instantons and remove the anisotropy.  If this turns out to be the case, then to produce a splitting between soft terms one must somehow modify the construction so that it is consistent with both anisotropy-generating poly-instantons and the wrapping of a small volume cycle by a stack of D7 branes. A fully consistent top-down construction of such anisotropic models would be welcome.

\acknowledgments{JC is funded by the Royal Society with a University Research Fellowship, and by the European Research Council under the Starting Grant `Supersymmetry Breaking in String Theory'. SA is funded by an STFC studentship.}

\end{document}